\documentclass[journal=apchd5,communication=article]{achemso}

\usepackage[version=3]{mhchem} 
\usepackage[T1]{fontenc}       
\usepackage{caption}
\usepackage{float}
\usepackage{geometry}
\usepackage{natbib}
\usepackage{setspace}
\usepackage{xkeyval}
\usepackage{subfig}
\usepackage{graphicx}

\author{Amir Youssefi}
\affiliation{Department of Physics, Sharif University of Technology, Tehran 11155-9161, Iran}
\alsoaffiliation{Department of Electrical Engineering, Sharif University of Technology, Tehran 11555-4363, Iran}
\author{Farzad Zangeneh-Nejad}
\affiliation{Department of Electrical Engineering, Sharif University of Technology, Tehran 11555-4363, Iran}
\author{Sajjad AbdollahRamezani}
\affiliation{Department of Electrical Engineering, Sharif University of Technology, Tehran 11555-4363, Iran}
\author{Amin Khavasi}
\affiliation{Department of Electrical Engineering, Sharif University of Technology, Tehran 11555-4363, Iran}
\email{khavasi@sharif.edu}

\title[An \textsf{achemso} demo]
{Analog computing by Brewster effect}


\begin{document}


\begin{abstract}
Optical computing has emerged as a promising candidate for real-time and parallel continuous data processing. Motivated by recent progresses in metamaterial-based analog computing [Science 343, 160 (2014)], we theoretically investigate realization of two-dimensional complex mathematical operations using rotated configurations, recently reported in [Opt. Lett. 39, 1278 (2014)]. Breaking the reflection symmetry, such configurations could realize both even and odd Green's functions associated with spatial operators. Based on such appealing theory and by using Brewster effect, we demonstrate realization of a first-order differentiator. Such efficient wave-based computation method not only circumvents the major potential drawbacks of metamaterials, but also offers the most compact possible device compared to the conventional bulky lens-based optical signal and data processors. 
\end{abstract}



Although analog computation has almost been disregarded by emergence of digital computation over years, it has continued to be used in some specialized applications to the present time. For example, in most cases, natural computation is analog, either because it benefits from continuous natural processes or exploits discrete but stochastic processes \cite{maclennan2007review,miller2014evolution}. In addition, theoretical results have shown that analog computation can escape from some restrictions of digital computation such as data conversion loss \cite{solli2015analog}. Therefore, analog computation is still known as an important alternative technology for digital computation.

To perform analog computation several different approaches have been investigated. These approaches can basically be classified into two categories, namely temporal and spatial analog computation. Although several valuable researches concerning temporal analog computation have been reported in recent years \cite{liu2008compact,slavik2008photonic,li2009arbitrary,park2010implementation,azaa2010ultrafast,li2011continuously,tan2013high,tan2013all,yang2014all,huang2015terahertz}, such proposals are complicated to be integrated due to their remarkably large size \cite{liu2008compact}. Therefore, in this Letter, we exclusively investigate the promoting spatial analog computation.

The idea of spatial computation conventionally appeared in analog computers, or ``calculating machines'' to make mechanical, electronic, and mechanical-electronic analog computers. However, this approach suffers from significant restrictions such as having relatively large size and slow response \cite{price1984history,clymer1993mechanical,preciado2008design,liu2008compact,slavik2006ultrafast,rivas2009experimental}.

To go beyond the aforementioned limitations of the spatial analog computation, two approaches have been investigated recently, called metasurface approach and Green's function (GF) approach \cite{silva2014performing,sihvola2014enabling}. The metasurface approach is based on the fact that the linear convolution $h(y)=f(y)*g(y)$ between an arbitrary electromagnetic field distribution $f(y)$ and the Green function $g(y)$ related to the desired operator of choice can be expressed in the spatial Fourier space as $H(k_{y})=F(k_{y})G(k_{y})$, in which $H(k_{y})$, $G(k_{y})$ and $F(k_{y})$ are the Fourier transform of their counterparts in the convolution equation \cite{abdollahramezani2015analog}. To this end, instead of implementing the Green's function $g(y)$ directly, a spatial Fourier transform is applied to the Green's function $g(y)$ to obtain the transformed Green's function $G(k_{y})$, and then by employing a gradient metasurface structure, $G(k_{y})$ is performed in the spatial Fourier domain. Therefore, in this method, the system should mainly include  three cascaded subblocks: 1) a Fourier transform subblock, 2) a properly adjusted metasurface spatial filter applying the $G(k_{y})$ operation in the spatial Fourier domain, and 3) an inverse Fourier transform subblock. Although this approach is applicable, it causes some fabrication complexity due to going into the spatial Fourier domain, and consequently, the need of two additional subblocks performing Fourier and inverse Fourier transforms \cite{farmahini2013metasurfaces,silva2014performing,pors2014analog,kou2016chip,zhang2016solving,chizari2016analog}. Moreover, making use of the mentioned additional subblocks implies an increase in the size of the system. Therefore, in this Letter, we will focus our attention to the second approach.

\begin{figure}
		\centering
		\includegraphics[trim=0cm 0cm 0cm 0cm,width=8.3cm,clip]{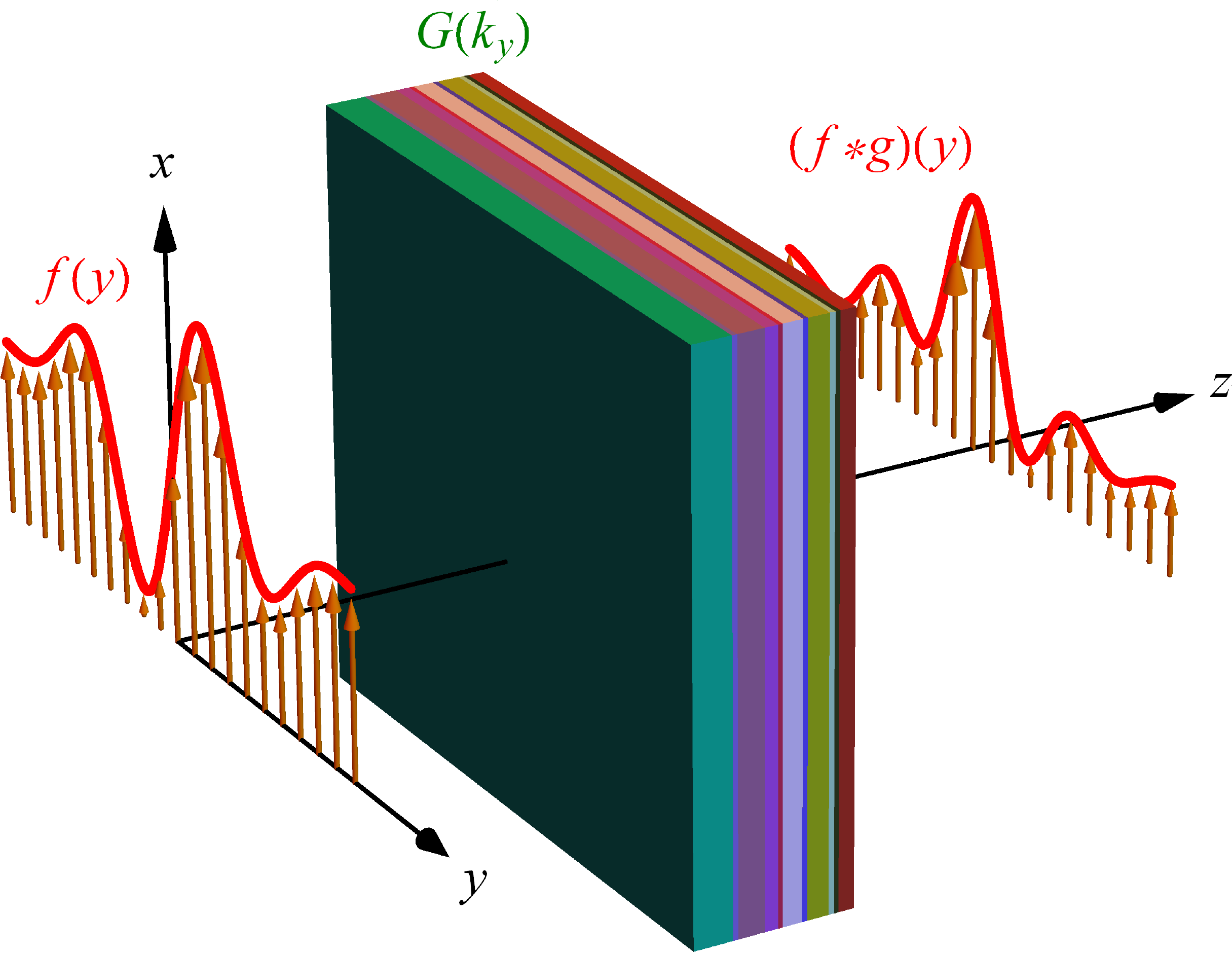}
		\captionsetup{justification=justified}
		\caption{Configuration of a multilayered slab performing Green's function of a desired mathematical operation. All layers are transversely homogeneous but can be longitudinally inhomogeneous.}\label{fig1}
	\end{figure}

In the second approach, namely Green's function method, mathematical computation is directly realized by suitably designing a multilayered slab which is homogeneous along the $x$ and $y$ axes as shown in Fig.~\ref{fig1}. The  multilayered metamaterial is designed such that it realizes an output function $h(y)$ regarding an input field distribution  $f(y)$, consistent with the  Green's function $g(y)$ associated with the operator of choice. Since in this approach the Green's function $g(y)$ is straightly performed, there is no need to go into the Fourier domain. Hence, one avoids the need of subblocks that perform Fourier and inverse Fourier transforms in the first method. As a result, not only can the fabrication complexity of the system be reduced, but also the whole structure will be miniaturized.

There are, however, two major drawbacks regarding the multilayered system proposed in \cite{silva2014performing}. First, only Green's functions with even symmetry in the spatial Fourier domain such as the operator of second-order derivative could be realized by means of the  multilayered structure shown in Fig.~\ref{fig1}. This is due to the reflection symmetry of the system \cite{silva2014performing}. However, there are many appealing and important Green's functions having odd symmetry in the Fourier domain such as first-order derivative and integration operators. The second problem is that the relative permittivity and the thickness of each layer of the multilayered slab are calculated using a fast synthesis approach based on the simplex optimization method which leads to non-practical values of relative permittivities and thicknesses. As already shown in \cite{doskolovich2014spatial}, odd Green's function can be realized by a rotated configuration or, in other words, by oblique incidence. This is achieved by breaking the reflection symmetry of the structure. In this contribution, our aim is to propose a scheme to realize the basic differentiator by just an interface using Brewster effect to tackle the second drawback.

\begin{figure}
		\subfloat[]{\includegraphics[width = 0.9\columnwidth]{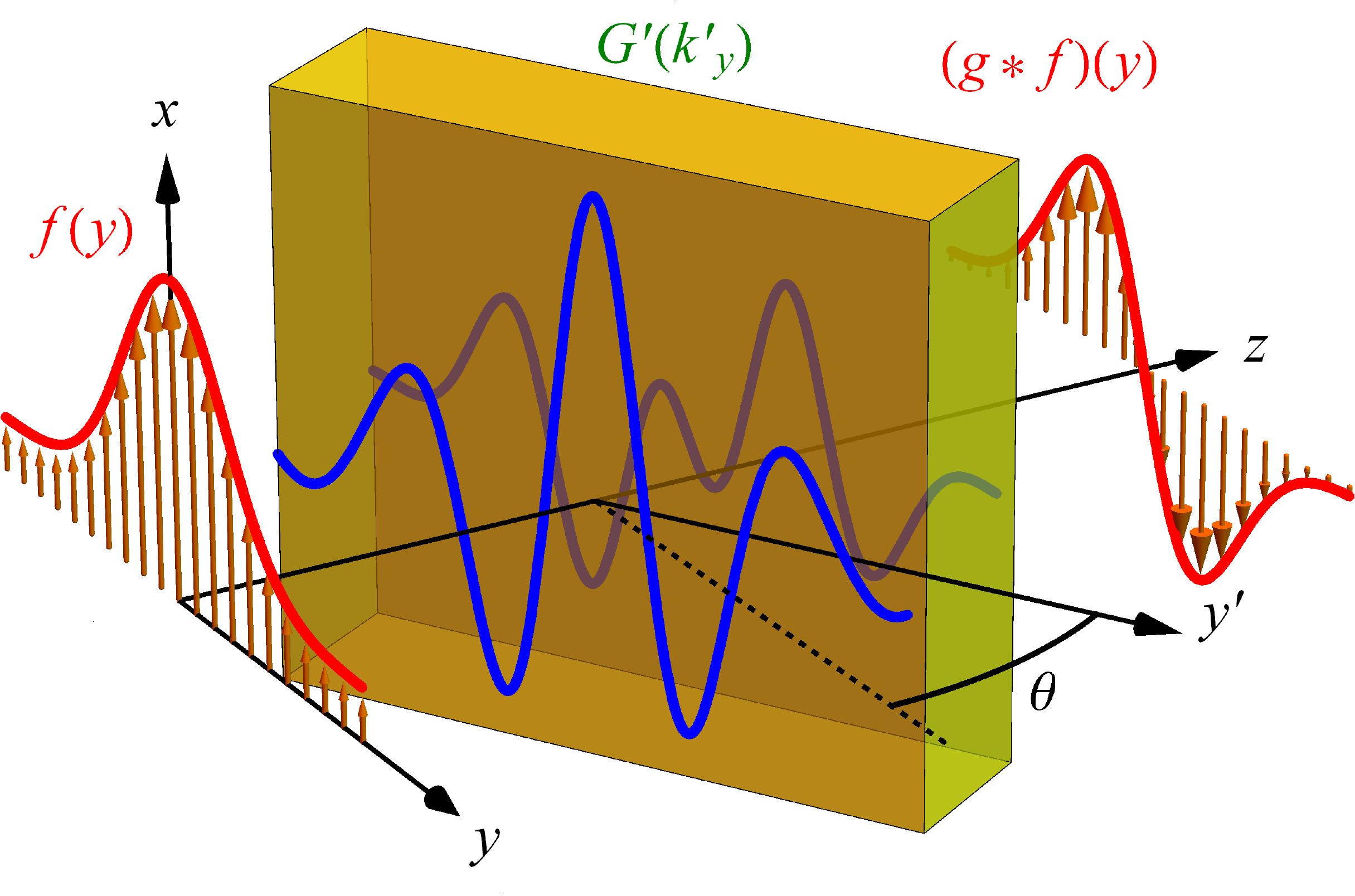}}\\
		\subfloat[]{\includegraphics[width = 0.9\columnwidth]{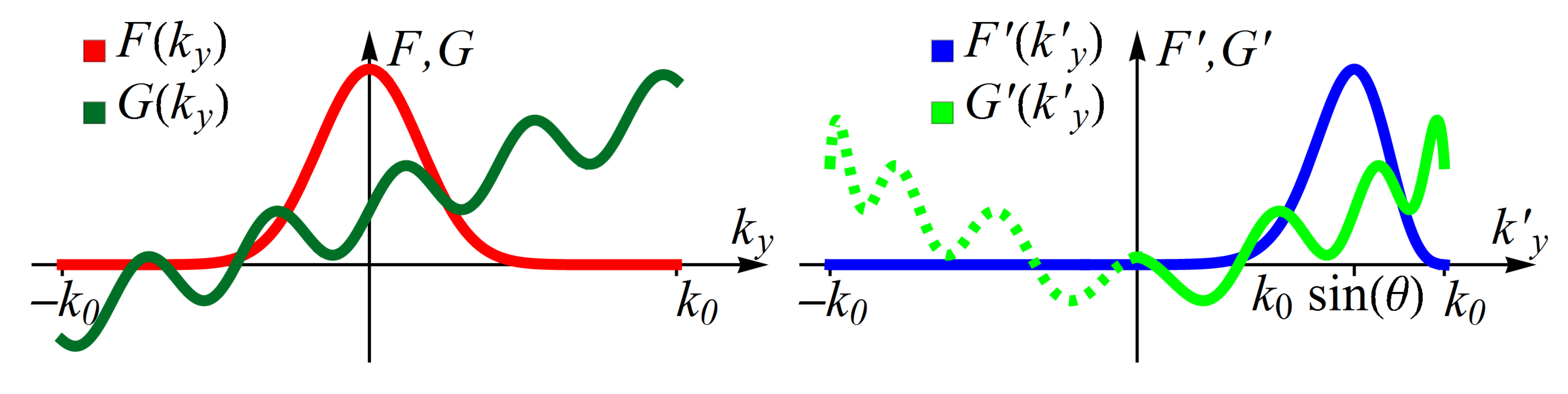}}
		\centering
		\captionsetup{justification=justified}
		\caption{(a) The proposed configuration with rotated angle $\theta$ for realizing an arbitrary even or odd Green's function. (b) Input signal $F^(k_{y})$, a defined Green's function $G(k_{y})$ (left), their corresponding transformation in the primmed coordinate (right).}\label{fig2}
	\end{figure}

The schematic of the system under study is shown in Fig.~\ref{fig2}(a). As it is seen, the reflection symmetry of the system is broken by employing a rotated structure instead of a straight one. An input wave characterized by field distribution of $f(y)\hat{x}$ with the bandwidth of $W$ in the spatial Fourier domain propagates along $z$ direction and incident onto the structure rotated by angle of $\theta$. The transformed Green's functions of the designed structure in the primed and unprimed coordinates are assumed to be $G(k_{y})$ and $G^{\prime}(k_{y}^{\prime})$, respectively, in which $k_{y}$ and $k_{y}^{\prime}$ are the Fourier variables in the unprimed and primed coordinates. Moreover, the detection direction is defined along the $y$-axis. Since the whole system has reflection symmetry in the primed coordinate system, the Green's function $G^{\prime}(k_{y}^{\prime})$ has to be even in the spatial Fourier domain. However, under certain circumstances, the Green's function $G(k_{y})$ associated with the unprimed coordinate can be considered odd.

To clarify the aforementioned circumstances, the relation between wave-numbers $k_{y}$ and $k_{y}^{\prime}$ associated with the primed and unprimed spatial Fourier domain has been expressed as follows
\begin{equation}
	k_{y}^{\prime}=k_{0}\textnormal{sin}(\theta+\textnormal{sin}^{-1}(\frac{k_{y}}{k_{0}}))
	\label{1}
\end{equation}
in which $k_{0}$ is the free space wave-number and $\theta$ is the rotated angle of the structure.
In this approach, the spatial spectrum of the input signal $f(y)$ is completely mapped to the right-half plane of primmed coordinate through the nonlinear transformation expressed in Eq.~\ref{1}. Using this equation and under the condition of $\textnormal{\textnormal{sin}}^{-1}({W}/{k_{0}})\leq \theta \leq \textnormal{\textnormal{cos}}^{-1}({W}/{k_{0}})$, the corresponding relationship between $G(k_{y})$ and $G^{\prime}(k_{y}^{\prime})$ is then obtained as
\begin{equation}\label{2}
	G^{\prime}(k_{y}^{\prime})=G(k_{0}\textnormal{sin}(\textnormal{sin}^{-1}(\frac{|k_{y}^{\prime}|}{k_{0}})-\theta))
\end{equation}
It is notable that under the applied restriction on the rotation angle $\theta$, not only is the overall spectrum of signal transformed to the right-half plane, but also it  remains in the span [$0,k_{0}$].
The approach to perform arbitrary Green's function $G(k_{y})$ is now obvious. First, one should apply the transformation given in Eq.~\ref{2} to the Green's function $G(k_{y})$ to obtain the associated Green's function $G^{\prime}(k_{y}^{\prime})$, and then implement $G^{\prime}(k_{y}^{\prime})$ in the primed coordinate. Accordingly, the effect of $G^{\prime}(k_{y}^{\prime})$ on $F^{\prime}(k_{y}^{\prime})$ is exactly the same as the effect of $G^(k_{y})$ on $G^(k_{y})$, as shown in Fig.~\ref{fig2}(b).

\begin{figure}
		\centering
		\captionsetup{justification=justified}
		\includegraphics[trim=0cm 0cm 0cm 0cm,width=8.3cm,clip]{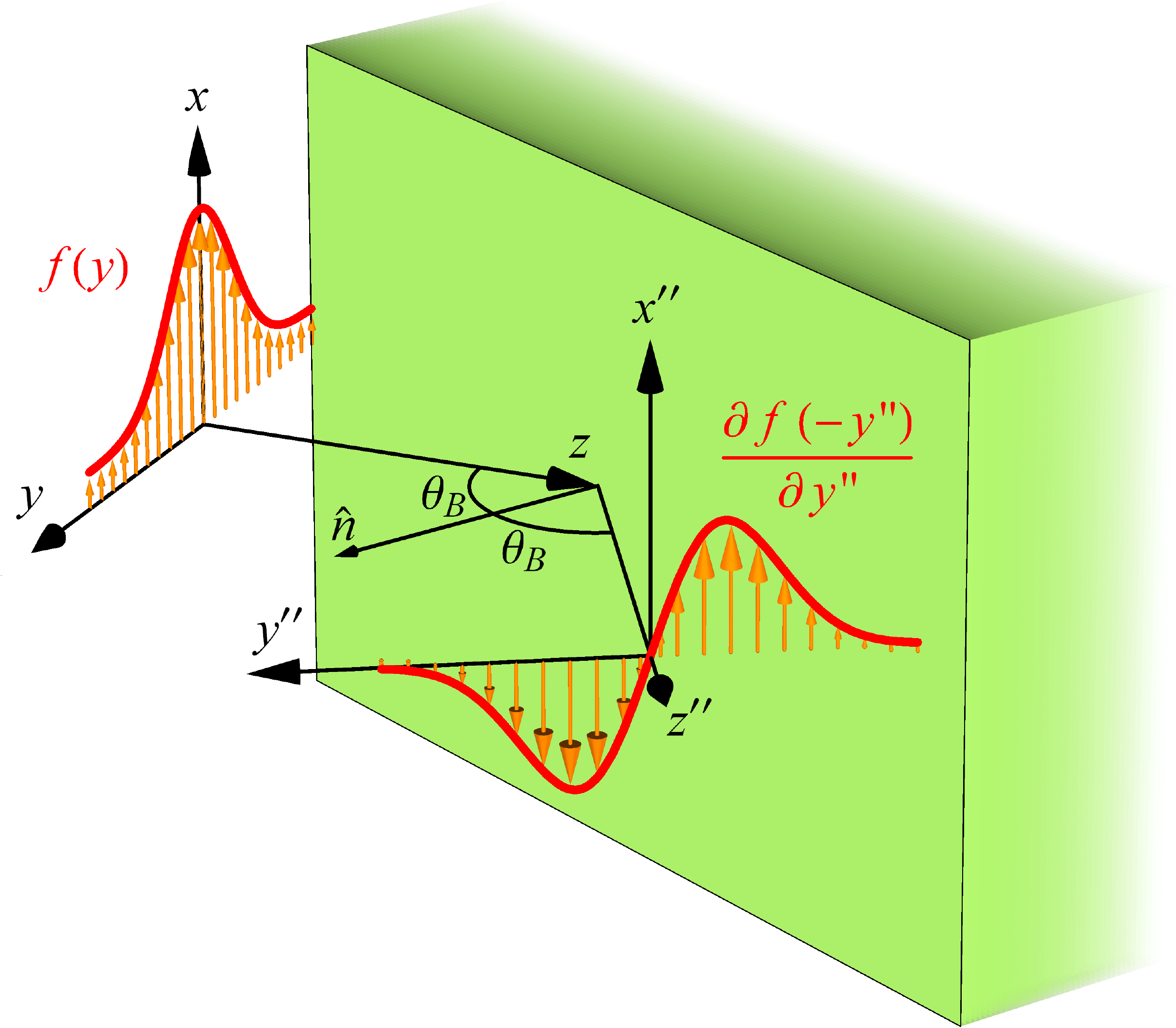}
		\captionsetup{justification=justified}
		\caption{Sketch of the system under study as well as perspective view of the signal waves incident onto and reflected from the interface at Brewster angle to realize first-order differentiation. Primed coordinate is orthogonal to the structure in which $z^{\prime}$ is parallel to $\hat{n}$. The reflected signal is received in $(x'',y'',z'')$ coordinates.}\label{fig3}
	\end{figure}

We further focus our attention to implement the Green's function of first-order derivative operators as an important example of odd Green's functions. Fig.~\ref{fig3} depicts our suggested structure to realize the Green's function of first-order derivative operator. As it is observed, our proposed structure is only composed of one interface. A beam with an arbitrary profile $f(y)$ is obliquely incident from free space onto the boundary of another medium with the refractive index $n$. According to Eq.~\ref{2}, since the Green's function $G(k_{y})$ of the first-order derivative operator is zero at $k_{y}=0$, the corresponding Green's function $G^{\prime}(k_{y}^{\prime})$ in the primed coordinate should be zero at $k_{y}^{\prime}=k_{0}\textnormal{sin}(\theta)$. Since neither TE nor TM polarized waves have zero transmission coefficient, it is not possible to obtain the first-order derivative of the incident wave on the transmission side. However, it is well-known that the reflection coefficient of an incident TM polarized wave becomes zero at the Brewster angle defined as
\begin{equation}
	\theta_{B}=\textnormal{tan}^{-1}(n)
	\label{3}
\end{equation}
Therefore, to satisfy the aforementioned criteria for the Green's function $G^{\prime}(k_{y}^{\prime})$, we consider the case in which a TM polarized input field $f(y)$ obliquely incidents from air onto the boundary of the other medium at the Brewster angle, and choose the reflected field $h(y)$ to be the output of the system depicted in Fig.~\ref{fig3}.
Employing Taylor series expansion of  $G^{\prime}(k_{y}^{\prime})$, or equivalently the reflection coefficient, about the Brewster angle $\theta_{B}$, one obtains
{\setlength\arraycolsep{2pt}
	\begin{eqnarray}\label{4}
		\nonumber
		G(k_{y})=G^{\prime}(k_{0}\textnormal{sin}(\theta+\textnormal{sin}^{-1}(\frac{k_{y}}{k_{0}})))=~~~~~~~~~~ \\ \nonumber
		G^{\prime}(k_{0}\textnormal{sin}(\theta_{B}))+\frac{\partial G^{\prime}(k_{y}^{\prime})}{\partial k_{y}^{\prime}} \mid_{k_{0}\textnormal{sin}(\theta_{B})} \times k_{y}\textnormal{cos}(\theta_{B})+O(k_{y}^2) \\
		=-(\frac{n}{2}-\frac{1}{2n^{3}})\frac{k_{y}}{k_{0}}+O(k_{y}^2)~~~~~~~~~~~~~~~~
	\end{eqnarray}}where we have used the fact that $G^{\prime}(k_{0}\textnormal{sin}(\theta_{B}))=0$.

\begin{figure}
			\centering
			\includegraphics[trim=0cm 0cm 0cm 0cm,width=8.3cm,clip]{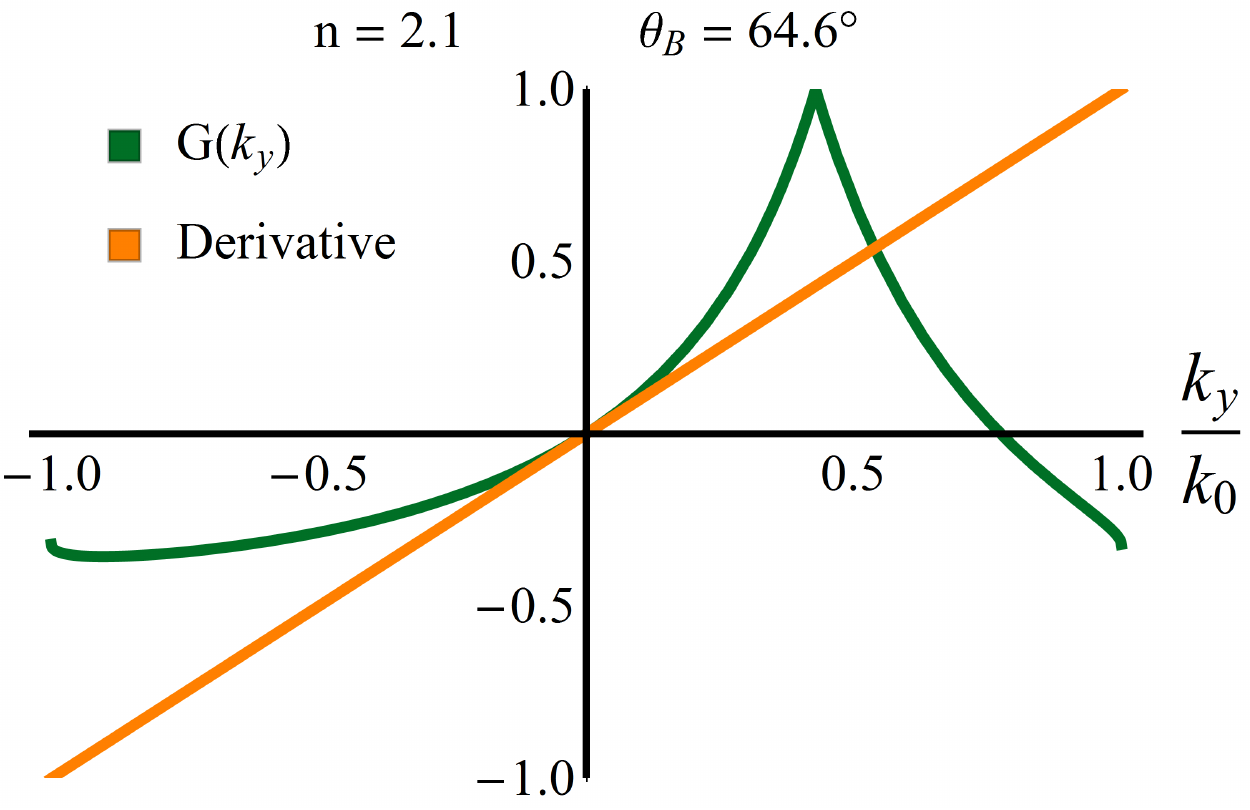}
			\caption{Distribution of the exact and approximated Green's function associated with the first-order derivative operation for $n=2.1$ and $\theta_{B}=64.6^{\circ}.$}
			\label{fig5}
		\end{figure}

\begin{figure}
			\centering
			\includegraphics[trim=0cm 0cm 0cm 0cm,width=8.3cm,clip]{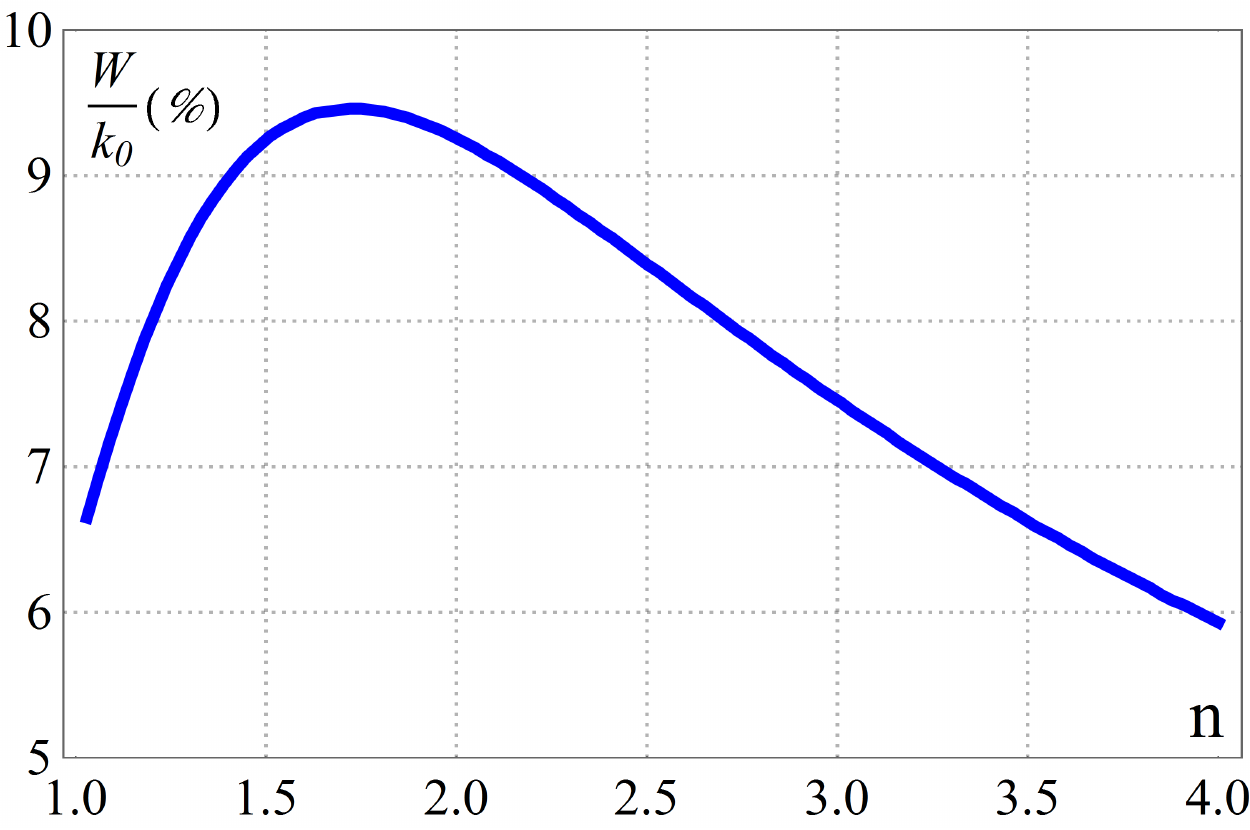}
			\captionsetup{justification=justified}
			\caption{Analytical result for the incident wave maximum spatial bandwidth versus refractive index of the secondary medium in the case of $10$\% error.}
			\label{fig4}
		\end{figure}

	It should be mentioned that the calculated Taylor series expansion in Eq.~\ref{4} is under the assumption that $k_{y}<<k_{0}$, or equivalently $ W<<k_{0}$. This assumption satisfies the restriction on the rotation angle $\theta$ applied in Eq.~\ref{2}. Fig.~\ref{fig5} depicts the exact Green's function $G(k_{y})$ and its approximation around $k_{y}=0$ based on Taylor series for $n=2.1$ and $\theta_{B}=64.6^{\circ}$. As it is observed in this figure, making use of the linear first order approximation of $G(k_{y})$ proposed in Eq. \ref{4}, its exact value can be estimated well around $ky=0$.\\
	\indent Using Eq.~\ref{4}, the corresponding operator to the Green's function $G(k_{y})$ can easily be  obtained as
	\begin{equation}\label{5}
		L[f(y)]\approx \frac{i}{k_{0}}(\frac{n}{2}-\frac{1}{2n^{3}})\frac{d}{dy}f(-y)
	\end{equation}
	which is the operator of the first-order derivative multiplied by a definite scale factor. We note that the intrinsic drawback of any differentiator is the low amplitude of its output, because the Green's function of the ideal differentiator is zero at $k_{y} = 0$. However, the proposed Brewster differentiator's efficiency is larger than $1$ (compared to the ideal differentiator) for $n>2.1$, according to Eq. \ref{5}. \\
	\indent It should be emphasized, there is no mechanism to control the width of the angular interval (around the zero reflection), where the reflection coefficient is able to approximate the Green's function of the ideal differentiator accurately. In order to obtain the maximum bandwidth $W$ in which the first two terms of Taylor series approximates the exact Green's function well, we set the maximum error of the estimation to be less than $10$\% based on the definition ${{e_{G}}=(\|G-G_{\textnormal{exact}}\|})/\|G_{\textnormal{exact}}\|$ \cite{le2008photonic}. The parameter $W$ versus the refractive index $n$ is plotted in Fig.~\ref{fig4}. Using this figure, we can readily determine the maximum bandwidth of the incident wave for a specific refractive index $n$.

\begin{figure}
			\includegraphics[trim=0cm 0cm 0cm 0cm,clip,width=8.3cm]{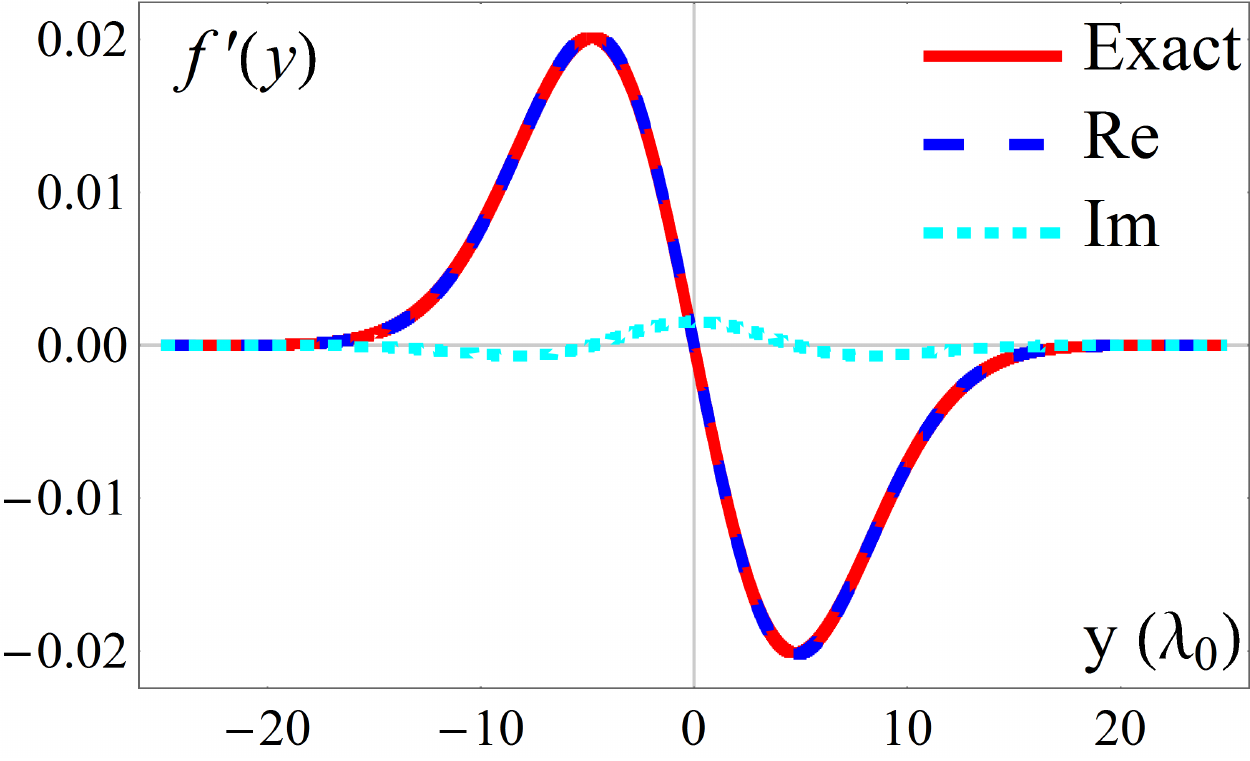}
			\centering
			\captionsetup{justification=justified}
			\caption{Analytical results for a signal wave including a Gaussian field with bandwidth of $W=0.1k_{0}$ and beamwidth $32\lambda_{0}$ incident onto the secondary medium depicted in Fig.~\ref{fig3}.}
			\label{fig7}
		\end{figure}

\begin{figure}
			\centering
			\includegraphics[trim=0cm 0cm 0cm 0cm,clip,width=8.3cm]{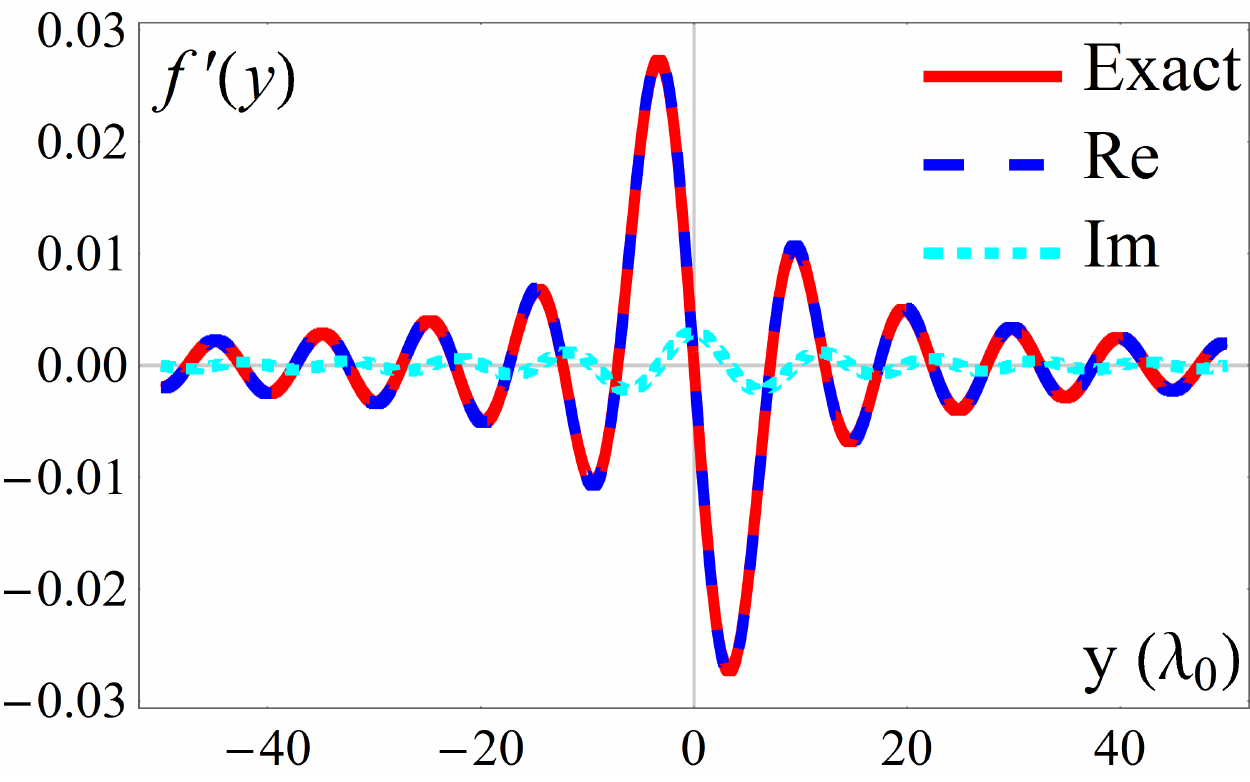}
			\caption{Analytical results for a signal wave including a Sinc function with bandwidth of $W=0.09k_{0}$ impinging on the secondary medium depicted in Fig.~\ref{fig3}.}
			\label{fig11}
		\end{figure}

	To evaluate performance of the proposed approach for realizing the Green's function of first-order derivative operator, we first investigate the case in which a Gaussian field $f(y)$ with the spatial bandwidth $W=0.1k_{0}$ and beamwidth $32\lambda_{0}$ impinges on the boundary of a medium with refractive index $n=2.1$ at the Brewster angle. 
	Fig.~\ref{fig7} illustrates the calculated first-order derivative of the incident wave obtained based on Brewster angle scheme compared with the exact analytical result. The results are in excellent agreement with $5$\% error according to formula ${{e_{f}}=(\|f^{\prime}-f^{\prime}_{\textnormal{exact}}\|})/\|f^{\prime}_{\textnormal{exact}}\|$.
	As another example, an electromagnetic wave which has  a Sinc function profile with the bandwidth of $W=0.09k_{0}$ is assumed to be incident onto the boundary of the same secondary medium at the Brewster angle. The obtained first-order derivative of the input field according to Brewster angle scheme as well as the exact result are  shown in Fig. \ref{fig11}. An error of $e_{f}=10~\%$ in the spatial domain has been calculated between theses two results. Since Sinc function in spatial domain 
	is mapped to a rectangular function with $W=0.09k_{0}$ in spectral space, we expect that spectral space error to be equal to spatial domain error, i.e., $e_{G}=e_{f}=10~\%$. Finally, we remark that higher order derivative operators such as the operator of second-order derivative can easily be  implemented by cascading two first-order derivative structures.

	In summary, stimulated by recent breakthrough in metamaterial-based mathematical operations \cite{silva2014performing}, we reviewed how a rotated structure can realize odd Green's functions by breaking the reflection symmetry \cite{doskolovich2014spatial,golovastikov2015spatial}. On the other hand, even Green's functions can be performed by conventional non-rotated structures \cite{bykov2014optical}. Based on such rotated configurations, a Brewster differentiator was realized to derivative spatial signals more accurately. High-order differentiators could readily be realized by a well-arranged array of the proposed first-order differentiator. Such appealing finding may lead to large improvements in image processing and analog computing.

\providecommand{\latin}[1]{#1}
\providecommand*\mcitethebibliography{\thebibliography}
\csname @ifundefined\endcsname{endmcitethebibliography}
  {\let\endmcitethebibliography\endthebibliography}{}

\newpage

\end{document}